\documentclass{jfm}

\usepackage{graphicx}
\usepackage{natbib}
 \usepackage{float}
 
\usepackage{color}
\usepackage{epstopdf}

\ifCUPmtlplainloaded \else
  \checkfont{eurm10}
  \iffontfound
    \IfFileExists{upmath.sty}
      {\typeout{^^JFound AMS Euler Roman fonts on the system,
                   using the 'upmath' package.^^J}%
       \usepackage{upmath}}
      {\typeout{^^JFound AMS Euler Roman fonts on the system, but you
                   dont seem to have the}%
       \typeout{'upmath' package installed. JFM.cls can take advantage
                 of these fonts,^^Jif you use 'upmath' package.^^J}%
       \providecommand\upi{\pi}%
      }
  \else
    \providecommand\upi{\pi}%
  \fi
\fi

\ifCUPmtlplainloaded \else
  \checkfont{msam10}
  \iffontfound
    \IfFileExists{amssymb.sty}
      {\typeout{^^JFound AMS Symbol fonts on the system, using the
                'amssymb' package.^^J}%
       \usepackage{amssymb}%

      }{}
  \fi
\fi

\ifCUPmtlplainloaded \else
  \IfFileExists{amsbsy.sty}
    {\typeout{^^JFound the 'amsbsy' package on the system, using it.^^J}%
     \usepackage{amsbsy}}
    {\providecommand\boldsymbol[1]{\mbox{\boldmath $##1$}}}
\fi

\newcommand\Pen{\mbox{\textit{Pe}}}  % Peclet number
\newsavebox{\astrutbox}
\sbox{\astrutbox}{\rule[-5pt]{0pt}{20pt}}

\newcommand\etc{etc.\ }
\newcommand\eg{e.g.\ }

\title{Near-wall nanovelocimetry based on Total Internal Reflection Fluorescence with continuous tracking}

\author[]%
{Zhenzhen Li $^{1*}$%
 ,\ns
 Lo\"ic D'eramo  $^{1*}$  \thanks{Email address for correspondence:  loic.deramo@espci.fr  $^*$\textit{both authors contributed equally}},
Choongyeop Lee $^1$,
  Fabrice Monti $^1$,
  Marc Yonger $^1$,
  Patrick Tabeling $^1$\\
Benjamin Chollet $^2$, Bruno Bresson $^2$,
Yvette Tran $^2$\\}

\affiliation{$^1$ MMN, CNRS, ESPCI Paris-Tech, 10 rue Vauquelin, 75005 Paris, France\\
$^2$ PPMD-SIMM, CNRS, ESPCI Paris-Tech, 10 rue Vauquelin, 75005 Paris, France  \\ [\affilskip]  }

\pubyear{ }
\volume{ }
\pagerange{ }
\date{?; revised ?; accepted ?. - To be entered by editorial office}

\begin{document}

\maketitle

\begin{abstract}
The goal of this work is to make progress in the domain of near-wall velocimetry. The technique we use is based on the tracking of nanoparticles in an evanescent field, close to a wall, a technique called TIRF (Total Internal Reflection Fluorescence)-based velocimetry.   At variance with the methods developed in the literature, we permanently keep track of the light emitted by each particle during the time the measurements of their positions (\textquoteleft altitudes\textquoteright) and speeds are performed. A number of biases affect these measurements: Brownian motion, heterogeneities induced by the walls, statistical biases, photobleaching, polydispersivity and limited depth of field. Their impacts are quantified by carrying out Langevin stochastic simulations, in a way similar to \cite{breuer09}. By using  parameters calibrated separately or known, we obtain satisfactory agreement between experiments and simulations, concerning the intensity density distributions, velocity fluctuation distributions, and the slopes of the linear velocity profiles. Slip lengths measurements, taken as benchmarks for analysing the performances of the technique, are carried out by extrapolating the corrected velocity profiles down to the origin along with determining the wall position with an unprecedented accuracy.   For hydrophilic surfaces, we obtain 1$~\pm~$5~nm for the slip length in sucrose solutions, and 9$~\pm~10$ nm in water, and for hydrophobic surfaces, 32$~\pm~5$~nm for sucrose solutions and 55$~\pm~$9 nm for water. The errors (based on 95\% confidence intervals) are significantly smaller than the state-of-the-art, but more importantly, the method demonstrates for the first  time a capacity to measure slippage with a satisfactory accuracy, while providing a local information on the flow structure with a nanometric resolution. Our study confirms the discrepancy already pointed out in the literature between numerical and experimental slip length estimates. With the progress conveyed by the present work, TIRF based technique with continuous tracking can be considered as a quantitative method for investigating flow properties close to walls, providing both global and local information on the flow.

\end{abstract}

\newpage

\section{Introduction}

\paragraph{} Velocimetry techniques based on Total Internal Reflection Fluorescence (TIRF) have enlightened our understanding of the behaviour of Newtonian flows and particles near boundaries. The method was  pioneered by Yoda \citep{yoda03,yoda04,yoda05,yoda06,yoda07,yoda08,yoda10} and Breuer \citep{breuer04,breuer06,breuer06q,breuer07e,breuer09,breuer09-2}, and further developed or used by several authors \citep{kazoe,bouzigues}.  It consists in seeding the fluid with fluorescent nanoparticles and operate in an evanescent field near a wall/liquid interface. In such conditions, the fluorescent particles emit a light whose intensity is expected to decrease exponentially with their altitudes (i.e. their distances to the wall) and it becomes envisageable, by translating intensities into distances, to determine the location of each of them with respect to the wall without being subjected to diffraction limit. With the cameras available today on the market, the theoretical resolution of the technique is subnanometric, but in practice, for a number of reasons that will be explained later, the best resolution reported thus far in the literature is  30 nm \citep{yoda10}. Still this represents an improvement by more than one order of magnitude in comparison with well resolved techniques such as $\mu$PIV (micro Particle Image Velocimetry) \citep{santiago98, santiago99, joseph, tretheway02, zhanhuali}.

Reaching nanometric resolutions on the local measurement of velocity, diffusion constant, speed distributions, \etc is a breakthrough for flow instrumentation. It opens the possibility to analyse nanoflows in a great variety of contexts \citep{hu-guoqing, sparreboom, mijatovic}. Examples concern lubricating films in concentrated emulsions and foams \citep{kimura, schmid, briceno}, depleted layers in polymer solutions \citep{PG-DeGennes-macro,vincent90}, grain dynamics in microgel concentrated suspensions \citep{sessoms, meeker}, Debye layers\citep{bouzigues}, grafted brush structure \citep{PG-DeGennes-macro80, murat}, polymer melt near-wall behaviour \citep{brazhnik, brochard},  \etc). TIRF based velocimetry has thereby the potential to open new interesting avenues in fluid dynamics research. 

\paragraph{}Nonetheless, TIRF based velocimetry suffers from a number of limitations and artefacts that have been analysed by a number of investigators \citep{breuer06,breuer07e,breuer03,yoda05,yoda07,yoda10}.  The main problems are the followings:

\begin{itemize}

\item {\it Brownian wandering }
Nanoparticles are subjected to Brownian motion while the exposure times, or the delay times between two successive captures are limited  by the performances of the camera. During the exposure time $\tau$ (typically, one or two ms)  particles explore a region equal $\sqrt{2D\tau}$, in which $D$ is the diffusion coefficient of the particle in the fluid. Using Einstein estimate of the diffusion constant, one finds that there is an optimal scale $l_{opt}$, defined by 
\begin{equation}
l_{opt}=\left(\frac{kT\tau}{3\pi\mu}\right)^{1/3}
\label{loptimal}
\end{equation}
for which particle radius equals to Brownian standard type deviation (here, k is the Boltzmann constant, $T $ the absolute temperature, $\mu$ the fluid viscosity). Particles much smaller than $l_{opt}$ develop large Brownian excursions and therefore probe large volumes. Particles larger than $l_{opt}$ also degrade the resolution because of their size. Therefore, expression \ref{loptimal} provides a reference scale for establishing the spatial resolution that TIRF based velocimetry can achieve. It turns out that the best resolution achieved to-day (30 nm) 
coincides, in terms of order of magnitude, with $l_{opt}$.  In the future, progress on the spatial resolution of TIRF based velocimetry will obviously be facilitated by progress in camera technology, but to-day the possibility of improving the situation with the existing cameras is worth being considered.

\item {\it Biases induced by the heterogeneities of the spatial distribution of the particles.}
The distribution of particles is not homogeneous in space, which generates difficulties in the interpretation of the averaged measurements. In aqueous solutions, and in typical situations, the colloidal particles seeding the flow are negatively charged and the wall develops negative surface charges. A Debye layer builds up, repelling the colloids and thus depleting the near wall region in particles  \citep{oberholzer}. Dielectrophoretic effects, pointed out recently by Yoda \citep{yoda14} may also contribute to shape the heterogeneity field. Other sources of inhomogeneities in the spatial distributions of the particles are diffusion gradients, inducing a drift oriented away from the wall, and closer to the walls, short range forces \citep{koch}. Since all measurements involve spatial averaging, either for statistical reasons, or because of Brownian wandering, the interpretation of the averages obtained with such heterogeneities is delicate. In the present state of the art, Yoda \citep{yoda10, yoda06} partitioned the space into three regions, assuming homogeneity in each of them. The velocity profiles obtained with this approach, being composed, for reasons linked to Brownian motion, of only three points in the 200-500 nm range \citep{yoda10}, are modestly resolved. One question is whether it is possible to improve the situation, for instance by reducing the size of the intervals over which averages are determined, while ensuring acceptable statistical convergence and controlling or reducing statistical biases.  
\item {\it Polydispersity of the particle characteristics.}
The particles currently used in the TIRF velocimetry experiments have polydispersivities in size, quantum efficiency and numbers of incorporated fluorophores. Since the determination of the position relies on intensity measurements, this variability impacts  the accuracy at which the particles can be localized. This difficulty can in principle be eliminated by carrying out statistical averaging, provided that the heterogeneity issue raised above can be addressed. In practice, using 100-200 nm fluorescent particles, for which polydispersivity lies between 5 and 10\% is acceptable, while using 20 nm, with polydispersitivities between 20-30\% is problematic.
\item {\it Position of the wall.}
Determining the wall location accurately is critical, since errors made at this level impact directly the accuracy of  the slip length determination. Determining the wall location necessitates separate measurements that must be made in situ, in order to keep the flow geometry and the instrumentation environment unchanged. One of the techniques consists in enhancing adsorption by adding salt (and then suppressing electrical screening), and measuring the intensity distribution of the particles adsorbed onto the wall  \citep{yoda10}. In the present state-of-the-art, these measurements are subjected to significant uncertainties, due, mostly, to an inaccurate analysis of the bleaching process.

\item {\it Relating particle speeds to flow speeds.}
Even though the particle positions and speeds were accurately determined, the  information would be not sufficient for determining the flow. Close to a wall, owing to various phenomena well documented in the literature (hindered diffusion, slowing down,...) the particle speed is not equal to the flow speed. Models must therefore be developed to convert particle speed data into flow speed data.  

\end{itemize}

\vskip0.1cm

\paragraph{} To-day, TIRF velocimetry has enlightened our understanding of the behaviour of Newtonian fluids close to a wall by establishing the structure of the velocity profiles and investigating how particles are transported. However, the biases and artefacts discussed above severely affect its capability to discuss slippage phenomena. The quantity characterizing slippage is the slip length, i.e., for TIRF based velocimetry, the length obtained by extrapolating the velocity profiles down to zero (thus the \textquotedblleft extrapolated slip length\textquotedblright). Slip length measurements have attracted the interest of a community for more than one decade and it is now well documented for the case of Newtonian fluids \citep{bocquet07, bocquet10}. Slip lengths can be taken to benchmark different techniques. At the moment, the best accuracy (based on standard type deviation) obtained on the slip length measurements by TIRF is $\pm$ 30 nm \citep{yoda10, breuer06}. This limited accuracy ranks TIRF velocimetry well below SFA \citep{charlaix05}, which currently achieves $\pm$ 2 nm, and other methods, such as pressure drop measurements \citep{breuer03} which achieves $\pm$ 5 nm. Worse, the actual TIRF-based slippage measurements do not allow to distinguish between hydrophilic and hydrophobic surfaces \citep{yoda10}; in another work \citep{breuer06}  slip lengths lying between 26 and 57 nm are found for hydrophilic surfaces, which disagrees with the generally accepted view that there is no slip on such surfaces. By failing to demonstrate a capacity to measure slip lengths with a satisfactory accuracy, one must admit that, at variance with most velocimetry techniques, TIRF-based nanovelocimetry cannot be envisioned yet as a quantitative tool for exploring flows in near-wall regions.

\paragraph{}The objective here is to make progress on TIRF based nanovelocimetry. By using a new methodology, based on continuous tracking of the particles, along with improving the precision of some determinations, we succeeded to significantly improve the performances of TIRF based nanovelocimetry in terms of spatial resolution and measurement accuracy.  In a nutshell, coupled to the accurate determination of the wall position, along with Langevin simulations for estimating the systematic biases, we reached a $\pm$ 5 nm on the slip  length measurements in sucrose solutions and a $\pm$ 10 nm in water (based on 95\% confidence intervals). This represents a substantial improvement compared to the state-of-the-art \citep{yoda10}. With these improvements,  we demonstrate for the first time that TIRF based velocimetry is an outstanding {\it quantitative} tool for exploring flow behaviors in the first hundreds nanometers near a wall.

\section{Description of the experimental set-up}

\begin{figure}
\centering
\includegraphics[width=120mm]{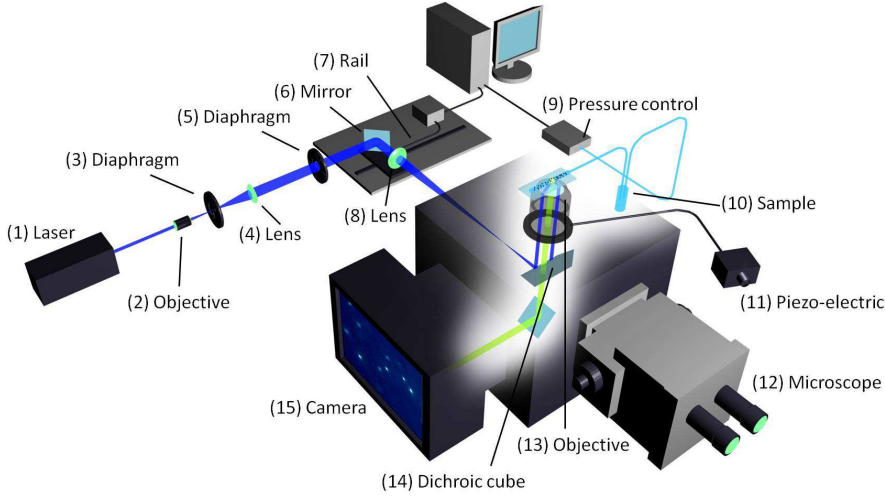}
\caption{Scheme of the TIRF setup. A laser beam is collimated through a high numerical aperture objective (100X, NA=1.46) with an incidence angle higher than the critical angle for total reflection. Laser is focused in the back focal plane of the objective with an achromatic doublet (focal length=200 mm) and the incidence angle is tunable by displacing a mirror in the optical path. The evanescent wave created within the channel restricts fluorescence to tracers (dots over the glass slide) in the vicinity of the surface, which is collected by the high-speed camera.} 
\label{setup}
\end{figure}

\paragraph{}The illumination system is sketched in figure \ref{setup}. A laser beam $(1)$ is initiated from a Sapphire laser (Coherent Sapphire 488-50) of wavelength 488 nm at output power 350 mW. This paralleled beam goes through an objective with 10X magnification $(2)$ to be focused on the focal plane. A diaphragm $(3)$ is placed on the same focal plane with 10 $\mu$m of diameter to let pass only the focused light. A lens $(4)$ with focal length 150 mm is placed behind the orifice, with its focal plane superposed with that of the 10X objective. The beam comes out from this lens being paralleled again but with an enlarged diameter at 2 cm. The implementation of $(2)(3)(4)$ is for producing perfectly paralleled laser beam with enlarged diameter, and eliminating lights due to unparalleled incidence and diffraction. The paralleled beam goes through a diaphragm $(5)$, which is fixed on the table during the first step of alignment of the laser, and serves as a reference point of the beam. The component $(6)$ is a mirror which reflects the beam at 90$^{\circ}$ and incident onto a lens $(8)$ to be focused on its focal plane; this focal plane is shared by the lens $(8)$ and the objective of the microscope. There is a dichroic filter cube (510 nm) situated between the lens $(8)$ and the objective, in order to reflect the incident laser beam up to the objective. The component $(6)$ and $(8)$ are fixed on a rail in order to be moved together horizontally without relative displacement, the rail is regulated by a home made Labview program. The fact of moving components $(6)$ and $(8)$ horizontally does not impede the laser beam to be focused on the focal plane of the objective, but results in a displacement of focusing point away from the optical line of the objective, which induces an angle of incidence different from 90$^{\circ}$ relatively to the micro channel.

\paragraph{}The advantage of using an objective for TIRF measurement relies in the convenience of sending incident light at a supercritical angle without a prism, and retrieving fluorescence information at the same time (see \cite{selvin_single-molecule_2008}). The objective has magnification 100X and a high numerical aperture at 1.46. The numerical aperture is expressed by the following equation:
\begin{equation}
\mathrm{NA}=n_i \sin(\Theta)
\end{equation}
where $n_i$ is the refractive index of the working medium of the objective, i.e. immersion oil ($n_i=1.518$). $\Theta$ is the half-angle of the maximum light cone ($\theta = 74.1^{\circ}$). The total reflection critical angle at the interface of water and glass slide is about 61.4$^{\circ}$ with the refractive index of water at 1.3327; and for the interface of sucrose solution at 40w\% and glass slide it is 67.1$^{\circ}$ with the refractive index of sucrose solution at 1.3981. The incidence angle is measured by a glass hemisphere, which projects the light spot onto millimetre paper. This angle $\theta$ is: 
\begin{equation}
\theta=\arctan(H/L)
 \end{equation}
where $L$ is the horizontal distance between the objective and the millimetre paper board. The light spot height $H$ is measured with a 1 mm precision, which corresponds to an error of 1 nm on the penetration length, i.e. less than 1\% error on its determination. 

\paragraph{}Another advantage of using objective with large numerical aperture and oil immersion consists in the increasing number of orders of diffraction collected by the lens, which reduces the size of the Airy diffraction disk, and increases the lens resolution. 

\begin{figure}
\centering
\includegraphics[width=140mm]{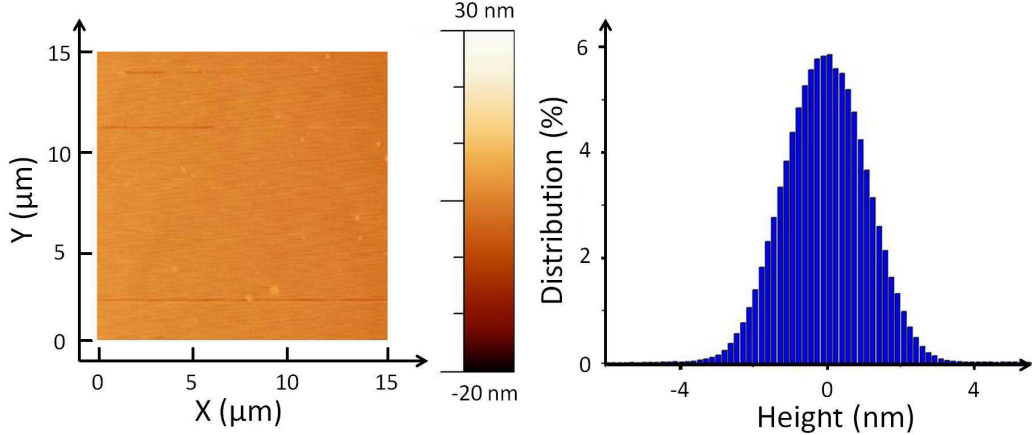}
\caption{State of hydrophobic surface coated on glass slide by n-octadecyltrichlorosilane, thickness measured by AFM (left), and RMS (right) indicating the roughness distributes around $0 \pm 1.3$~ nm at half-height.} 
\label{phobe}
\end{figure}

The expression of the evanescent field we apply reads, within the liquid:
\begin{equation}
I=I_{0}  \exp{(-z/p)}
\label{evanescentfield}
\end{equation}
in which $z$ is the distance from the glass/liquid interface (i.e. the channel wall), $p$ is the field depth of penetration and $I_{0}$ is the intensity of the evanescent field at $z=0$.
We use an Andor Neo sCMOS camera for the acquisition of data. It has several advantages. Firstly, it is able to work with a good sensitivity at 16 bit, which means 65 536 grey levels accessible, and so allows the distinction of two slightly different intensities. 
Secondly, by using the \textit{rolling shutter mode}, we can benefit from this lower noise and acquire data at high frame rate (400 fps), as 512*512 pixels pictures are taken during the experiments, with a 2.5 ms exposure time. Operated in \textit{overlay} mode, there is no time gap between two successive exposures, so that no information is lost. 
Thirdly, the camera provides 4GB of on-head image buffer, which overcomes the limitation of frame rate due to the eventually low write rate of hardware, and enables frame rates up to 400 images per second. 
Fourthly, the camera offers a FPGA generated hardware timestamp with the precision of 25 ns, i.e. 1\% of the exposure time, which has a negligible effect on the velocity calculation. 
Lastly, this sCMOS camera has an amplifier for each pixel, so that there might be a slight difference of amplification rate. However, the difference of amplification rate is 0.1\% in intensity, which induces 0.1\% error of altitude calculation, corresponding to 0.1 nm, which is a negligible effect in the experiments.  
The fluids are injected with a MFCS-FLEX pressure controller by Fluigent. The maximum pressure is 1000 mbar, with 1 mbar precision (0.1\% of full scale). The minimum pressure applied in water experiment is 20 mbar, which means that the source of error due to the pressure controller introduces an error on the stress calculation of maximum 5\% for water, and 1\% for 40w\% sucrose solution, inducing a maximal error on the deduced viscosity of 5\% and 1\% respectively.
DI water (DIW) and 40w\% of sucrose solution in DIW are studied. Fluospheres particles from Invitrogen (carboxylate modified 100 nm Yellow-Green Fluorescent particles F8803, solid 2w\%, $3.6\times10^{13}$ particles/ml) are used as tracers. Their peak excitation and emission wavelengths are 505 nm and 515 nm. The particle concentration seeded in 40w\% sucrose solution and DIW are respectively 0.0004w\% and 0.002w\% solid, so as to have up to 10 particles detected in each frame. 
The channel dimension is $L = 8.8$ cm in length, $w = 183~\mu$m in width and  $h = 18.3~\mu$m in height. With these dimensions, the hydrodynamic resistance has the right order of magnitude to operate with flow speeds well adapted to our particle tracking system.  The channel is designed in a coil shape in order to save space. With a $Re$ on the order of  $10^{-3}$, inertia is neglected and the small turning points of the channel does not induce any recirculation nor instabilities. Both hydrophilic and hydrophobic surfaces are studied. Hydrophilic surfaces are prepared by treating glass slides with Piranha solution (30v/v\% hydrogen peroxide and 70v/v\% of sulfuric acid) heated at 300$^\circ$ C until degassing ends. The obtained slides are bonded to PDMS channels with O$_2$ plasma treatment. Hydrophobic surfaces are obtained by coating a layer of n-octadecyltrichlorosilane on a slide treated by Piranha solution \citep{mcgovern}. The coated hydrophobic layer is 20 \AA ~thick according to ellipsometry measurements. The contact angle of a DIW drop on the coated layer is $105^\circ$, measured by a surface tension determination setup KR\"{U}SS DSA30. The roughness of the coated surface is measured by Atomic Force Microscope (figure \ref{phobe}): the relative roughness distributes around $0 \pm 1.3$ nm at half-height. The hydrophobic slide is combined with channel printed in NOA \citep{bartolo}. NOA-made channels are preferred to PDMS to avoid the alteration of the hydrophobic coating by the plasma exposition. \\

\setlength{\tabcolsep}{7pt}
\begin{table}
  \begin{center}
    \begin{tabular}{cccccc}
     {sample} &{$\dot{\gamma}(s^{-1})$}& {$V(\mu m/s)$}& {$Re$} & {$Pe$} &{$\mu(mPa.s)$}\\      
     \hline
     Sucrose $40\%$w & 200-600 & 560 - 1500 &$[2; 7]\cdot10^{-3}$ & 0.5-1.8 & 6.2\\
     Water & 200-800 & 560 - 1800 & $[10; 40]\cdot10^{-3}$ & 0.1-0.5 & 1\\
     \hline
    \end{tabular}    
  \end{center}     
\caption{Experimental conditions and non-dimensional numbers. $\dot{\gamma}$ the fluid shear rate. Reynolds number is defined as $Re=\rho V H/\mu$ and Peclet numbers defined as $Pe=\dot{\gamma}r_{0}^{2}/D_{0}$ \citep{breuer09}, where $V$ is the average velocity of fluid in the channel, $\rho$ density of the fluid, $H$ height of the channel, $D_{0}$ diffusion coefficient of particles in the fluids calculated by Stokes-Einstein equation, and $r_{0}$ the particle radius. Viscosity $\mu$ is measured by conventional rheometer at $20^{\circ}C$.}
\label{characteristic}
\end{table}

\section{Method of measurement of the particle locations and speeds}
\label{methodparticles}
During the acquisition, for each five seconds long experimental run, 2000 images are taken; and for each applied pressure, 6 runs are carried out to ensure a high enough number of detected particles around 80000, which will contribute to the statistics of the data analysis. The measurement of the intensity is the averaged intensity $\bar{I}$ over the exposure time of the camera $\tau$, with $\tau = 2.5$ ms. In this process, no particle is lost from the observation, since the two intervals of time are strictly consecutive. This represents a crucial difference with previous work (\eg \cite{breuer06}, \cite{yoda08}). The experimental conditions and characteristic numbers are summarized in table \ref{characteristic}. An interesting quantity that we used in the discussion of the results is the shear rate $\dot{\gamma}$  which is determined by the following formula:  
\begin{equation}
\dot{\gamma}\; = \frac{h \Delta P}{2 \mu L}
\label{relationglobale}
\end{equation}
where $\Delta P$ is the applied pressure and $\mu$ the fluid viscosity. \\

A custom made Matlab program is used to analyze the detected particles intensity and displacements. We use the tracking method developed by \cite{bonneau}. The subpixel $x$ and $y$ locations are obtained by convoluting the detected intensity spot with a 2-D Gaussian template, with standard deviation $\sigma\approx1.1$ pixels, and discretized on a support of size $5\times5$ pixels$^{2}$. The accuracy of coordinates determination is $0.1$ pixels with a signal-to-noise ratio greater than 10 \citep{bonneau}. This represents a spatial resolution of 6 nm. In this study of TIRFM, since the velocity close to the wall is small, and the time delay between two frames is short, the displacement of one particle from an image to the next one is much shorter than the distance between two detected particles, considering that the number of particles detected in one image is around 10 in a window of 33*33 $\mu$m. Thus an algorithm is implicated to identify the same particle which appeared in two successive images with a nearest neighbour search \citep{breuer09}, taking into account the estimated displacement and the diffusion effect. The technique is efficient, and mismatches are extremely rare events.\\
After particles identification, displacement between two successive frames can be calculated with a theoretical resolution of $6 \sqrt{2}\approx $ 8 nm. The time averaged velocity $\bar{V}_{X}$ and $\bar{V}_{Y}$ are calculated by displacement divided by time gap 2.5 ms. The time averaged intensity $\bar{I}$ is subtracted by camera background noise, and then normalized by the laser-only intensity at the pixel position. The apparent intensity of one identified particle $\bar{I}_{app}$ is the average between two successive frames:
$\bar{I}_{app}=(\bar{I}_{1}+\bar{I}_{2})/2$ where $\bar{I}_{1}$ and $\bar{I}_{2}$ are respectively the time averaged intensity measured in the first frame (called frame 1) and in the next one (called frame 2). In general, because of Brownian motion, the particle in frame 1 has not the same altitude as in frame 2. In order to check the effect of the altitude fluctuations on the results, we have compared velocity profiles constructed with a selection of particles travelling less than 50 nm in altitude between frames 1 and 2, with profiles obtained with particles  travelling up to 200 nm in altitude between the same frames. We  found that the two profiles are undistinguishable. We thus did not introduce any rule of selection of particles at this level.\\
The apparent altitude $\bar{z}_{app}$ of the particle is inferred by using the following expression derived from (\ref{evanescentfield}):
\begin{equation}
\bar{z}_{app}=p \log \frac{I_0}{\bar{I}_{app}}
\label{zappexperiment}
\end{equation}
in which $I_0$ is the intensity emitted by a particle located at the wall, a crucial parameter deserving a separate measurement (see section \ref{detI0}). 

The above formula allows to estimate the measurement accuracy of the apparent altitude and the resolution, i.e the  capacity to distinguish between the altitudes of two strictly identical particles. Error estimates  on $I_0$ (see below) and $p$ being on the order of 2 nm,  ${z}_{app}$ is determined with a similar accuracy. On the other hand, with the camera we use and with a 10:1 signal to noise ratio, the altitude resolution we obtain is an amazing $10^{-1}$ nm. In practice however, the resolution is limited by statistical constraints (see below).  

After these measurements are carried out for each individual particle, we proceed to statistical averaging over many particles. First, all the particles are sorted out according to their apparent altitudes $\bar{z}_{app}$. A box is an interval, along the $z$ axis, which contains particles whose apparent altitude belongs to the interval.The box sizes represents the spatial resolution at which the velocity profile is measured. Throughout the experiments, we take box sizes on the order of a few nanometres for distances larger than 200 nm from the wall. At less than 200 nm from the wall, box sizes can reach 50 nm. 
The reason is that we impose that each box includes  $N=1000$ particles, so as to guarantee statistical convergence. Thereby, since there are only a few particles very close to the wall, the box sizes must be increased in that region, at the expense of the resolution. \\
The statistically averaged velocities and apparent altitudes, calculated in each box, are determined by the relations:
\begin{equation}
<\bar{V}_X> \; = \frac{1}{N} \sum_{N} \bar{V}_X \hspace{20pt} ; \hspace{20pt} <\bar{V}_Y > \; = \frac{1}{N} \sum_{N} \bar{V}_Y \hspace{20pt}; <\bar{z}_{app}>= \frac{1}{N} \sum_{N} \bar{z}_{app} \hspace{20pt}
\end{equation}

in which $X$ and $Y$ are respectively the streamwise and cross-stream coordinates, and $\bar{V}_X$ and $\bar{V}_Y$ the corresponding time averaged speeds. The summations are carried out in each box over the 1000 particles belonging to it. The raw velocity profile is defined as $<\bar{V}_{X}>$ vs. $<\bar{z}_{app}>$.

\section{Langevin modeling of the particle trajectories and intensity measurements}
\label{langevinmod}
In this section, we describe the stochastic Langevin equations we used to interpret our measurements. We used the same approach as \cite{huang_direct_2006}. We consider here a population of spherical particles, with a mean radius $r_0$ and size dispersion $\sigma_r$, assuming a normal distribution for the radius $r$: 

\begin{equation}
p_R (r) = \frac{1}{\sigma_{r} \sqrt{2 \upi}}  \exp{-\frac{((r/r_0)-1)^2}{2 \sigma_{r} ^2}}
\end{equation}

These particles are transported by a pure no-slip shear flow $V(z)=\dot{\gamma} z$  (where $\dot{\gamma}$ is the shear rate) close to a wall located at z=0. 

The stochastic Langevin equations assign coordinates $x(t)$, $y(t)$, $z(t)$,  to each particle center at time $t$. The particles interact with the surface charges at the wall. We must introduce  zeta potentials $\zeta_p$ and $\zeta_w$ for the particles and the wall respectively to describe this effect. We neglect van der Waals forces, which act on scales much smaller than those we consider here (we checked that adding them does not affect the results of the paper).  For the sake of simplicity, we work with dimensionless quantities, using the following dimensionless variables $X = x/r_0$, $Y = y/r_0$ , $Z = z/r_0$ and $T = t /(r_0^2/D_0)$. Thus, $\delta T = \delta t /(r_0^2/D_0)$ is the dimensionless time increment taken for the integration of the trajectories. The discretized equations that describe the particle dynamics, originally written by \cite{ermak_brownian_1978}, are :

\begin{subeqnarray}
  X_{i+1} & = & X_i + F(Z_i)~ \Pen~ Z_i~ \delta T + S \left( 0,\sqrt{2 \beta_x (Z_i) \delta T } \right) \\[5pt]   \label{Langevin}
  Y_{i+1} & = & Y_i + S \left( 0,\sqrt{2 \beta_y (Z_i) \delta T } \right)  \\[5pt]
  Z_{i+1} & = & Z_i + \left.\frac{d \beta_z}{dZ} \right|  _{Z_i}\delta T + H(Z_i) \beta_z(Z_i) \delta T + S \left( 0,\sqrt{2 \beta_z(Z_i) \delta T} \right)  
\end{subeqnarray}
in which the streamwise ($\beta_x$ and $\beta_y$) and cross-stream ($\beta_z$) diffusion coefficients are defined by  :
\begin{subeqnarray}
\beta_x(Z) = &\beta_y(Z) =& 1- \frac{9}{16}Z^{-1} + \frac{1}{8}Z^{-3} - \frac{45}{256}Z^{-4} - \frac{1}{16}Z^{-5} + O(Z^{-6})  \\[5pt] 
&\beta_z(Z) =& \frac{6(Z-1)^2 + 2(Z-1)}{6(Z-1)^2 + 9(Z-1) + 2}
\end{subeqnarray}

The other dimensionless functions are defined by the following expressions:

\begin{equation}
F(Z_i) = \frac{U(z_i)}{z_i \dot{\gamma} }  \hspace{10pt} ; \hspace{10pt}
H(Z_i) = \frac{F_{pw}(Z_i) r_0}{k_B \Theta} \hspace{10pt} ; \hspace{10pt}
\Pen = \frac{\dot{\gamma}{r_0}^2}{D_0}
\label{velocityfactor}
\end{equation}

in which $U(z_i)$ is the fluid longitudinal velocity of a particle at the height $z_i$. 

Assuming that the wall particle interaction is described by DLVO theory, we obtain the following expression for $F_{pw}(Z_i)$ : 
\begin{equation} 
F^{el}_{pw} = 4 \pi \epsilon \frac{r_0}{\lambda_D} \left( \frac{k_B \Theta}{e} \right) ^2 \left(    \frac{\hat{\zeta_p} + 4 \gamma \Omega \kappa r_0}{1 + \Omega \kappa r_0}    \right) \left[�    4 \tanh \left( \frac{\hat{\zeta_w}}{4} \right)    \right]  \exp \left( -\frac{z-r_0}{\lambda_D} \right)
\label{electrostatic}
\end{equation}
where  
\begin{equation}
\hat{\zeta_p}  = \frac{\zeta_p e}{k_B \Theta} \hspace{10pt} ; \hspace{10pt}
\hat{\zeta_w}  = \frac{\zeta_w e}{k_B \Theta} \hspace{10pt} ; \hspace{10pt}
\gamma = \tanh \left( \frac{\hat{\zeta_p}}{4} \right) \hspace{10pt} ; \hspace{10pt}
\Omega = \frac{\hat{\zeta_p} - 4 \gamma}{2 \gamma ^3}
\end{equation}

and in which $k_B$ is the Boltzmann constant,  $\epsilon = \epsilon_w \epsilon_0$ is the dielectric constant of water,  $\zeta_p$ and $\zeta_w$ are zeta potentials of the particles and the wall respectively, and $\lambda_D$ is the  Debye length.  

The stochastic displacement of the particles due to Brownian motion is represented by the $S$ function, following a normal distribution with a zero mean value and the dimensionless diffusion length $L = \sqrt{ 2 \beta(Z) \delta T}$ as its standard deviation.

The total fluorescence intensity $I$ emitted by the particle at time $t_i$ must then be calculated.  Here, we consider that the particle is at some distance $z_f$ from the object plane of the objective. Bleaching is neglected, owing to the fact that the time of interest (2.5 ms) is much smaller than the fastest bleaching time (see~section~\ref{detI0}). The total intensity emitted by the particle, being proportional to the incident light intensity it collects, is thus given by the following expression: 

 \begin{equation}
I(Z)=I_0 \left(\frac{r}{r_0}\right)^3 \exp \left(-\frac{z}{p} \right) \frac{1}{1+\left( \frac{z-z_f}{P_{c}} \right)^2}
\label{intensityexpr}
\end{equation}

Where the parameters have the following definition:
\begin{itemize}
\item $I_0$ is the intensity emitted by a particle of radius $r_0$ located at the wall, i.e. at $z = r_0$, whose center is placed at the image plane of the camera sensor, i.e. at $z=z_f$.
\item $P_{c}$ is the depth of field of the optical system that images the particles \citep{joseph_direct_2005}. This parameter is estimated with the following formula (\ref{depthoffield}) :
 \end{itemize}
\begin{equation}
\label{depthoffield}
P_{c} = \frac{n \lambda_0}{\textrm{NA}^2 }+ \frac{na}{\textrm{NA}~M}
\label{Pc}
\end{equation}

 in which $n$ is the refractive index of the working medium $(n=1.518)$, $\lambda_0$ the excitation wavelength ($\lambda_0 = 488$ nm), $a$ the pixels size of the CMOS sensor ($a = 65$ nm), NA the numerical aperture of the objective ($\textrm{NA} = 1.46$) and $M$ the magnification (100X). The formula gives 348 nm for $P_{c}$.  $z_f$  represents the position of the image plane of the camera sensor, which is estimated to be 350 nm. The choice of this value is  supported by calibration experiments (see appendix \ref{calibration}). The effect of defocusing has not be considered thus far by the previous investigators, which seems justified by the fact that their objectives have smaller numerical apertures.

The procedure we use mimics rigorously the experimental methodology. It decomposes into the following steps:
\begin{itemize}
\item We generate particles whose initial positions span a range of altitudes between 100 nm and 700 nm, with uniform distributions. This interval corresponds to the range of altitudes within which particles are detected in the experiment. Outside this range, the particles are not detected either because of the existence of a depletion or the signal they emit is indistinguishable from the noise. 
\item Each particle wanders in space, and reaches some position (say, 1) after a time $\tau$ equal to the exposure time of the camera, i.e. 2.5 ms. This position is calculated by using equations \ref{Langevin}.  After another time $\tau$, the particle reaches a second position (say, 2) which is calculated in the same way.
\item For each particle, the intensity $I(Z)$ is temporally averaged between the initial time $t= 0$ and $ t=2 \tau$, the result being  $I_{app}$. A similar procedure applies for the altitude, leading to  $z_{mean}$ and for the horizontal coordinates of the particle.
\item The apparent altitude $z_{app}$ is calculated similarly as in equation \ref{zappexperiment}. The formula is:
 \begin{equation}
z_\mathrm{app} =r_0+ p \log\frac{I_0}{I_{app}}
\label{zappsimul}
\end{equation}
Note that $z_\mathrm{app}$ is generally different from $z_{mean}$.
\item The horizontal displacement vector of the particle between its mean positions 1 and 2 is calculated. The components of this vector are divided by  the separation time (i.e. $\tau$) to obtain the components of the apparent fluid speed, $V_{xapp}$ and $V_{yapp}$ as a function of $z_{mean}$. 
\item Boxes similar to the experiment are defined (see section \ref{methodparticles}). The quantities or interest (altitudes, speeds) are further averaged out statistically over thousands of particles located in each boxes.  Distributions are also obtained.\\

\end{itemize}
    
A final remark concerns the stationarity of our simulations. They are not stationary, since in the course of time, the particles tend to diffuse away. Nevertheless, the escape is extremely slow (its characteristic time is a fraction of seconds, i.e. several hundreds of times the time we consider here, $\tau$). The escape process is also slow with respect to the speed of establishment of the Debye layer, which takes only a fraction of $\tau$ to build up. Therefore, on the time scales we consider in the simulations, the leak of particles we evoke here can  be neglected.

\section{Using Langevin simulation to correct altitudes and speeds}

\begin{figure}
\centering
\includegraphics[width=120mm]{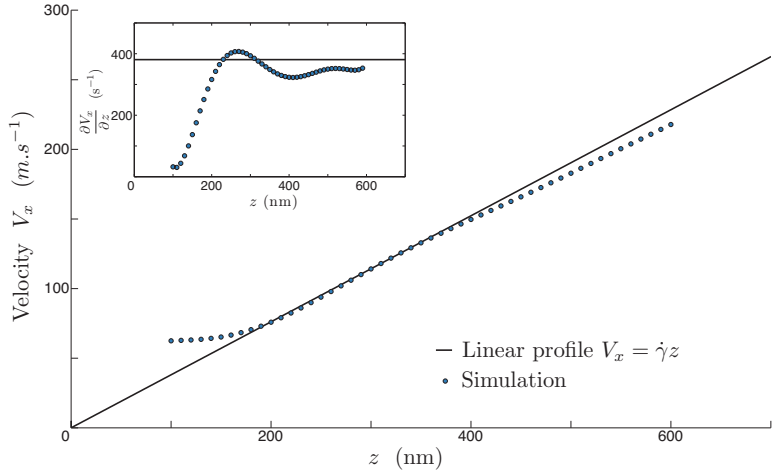}
\caption{Simulated velocity profiles of sucrose $40w\%$ solution on an hydrophilic surface for an inlet pressure $p = 195~$mbar. The parameters of the simlation are $\lambda_D = 22~$nm, $P_c = 348~$nm and $z_f = 300~$nm, and $p= 124$ nm. Inset shows the  local slope of the velocity profile.} 
\label{correction_simu}
\end{figure}

Now we make use of the simulations to discuss the relation between the apparent altitude $z_a$= $<z_\mathrm{app}>$ and the apparent longitudinal speed $V_a$=$<V_{xapp}>$ (where the brackets mean statistical averaging in the boxes).  In fact, each apparent quantity is a function of  $<z_{mean}>$ . Therefore, by eliminating  $<z_{mean}>$ , we can plot $V_a$ in function of  $z_a$, thus mimicking the experimental measurements we perform. The plot is shown in figure \ref{correction_simu}.

One sees that the apparent profile $V_a(z_a)$ departs from the true profile $V(z_a)=\dot{\gamma} z_a$. Close analysis shows that as a whole, the apparent profile is  shifted towards the right by a few nanometers, for reasons related to the coupling between Brownian fluctuations and convexity of the function $I(z)$. On the other hand, the lower part  of the apparent profile is curved upwards. This is due to the depletion in particles in the Debye layer. More explicitly, the explanation is the following: with the values we have taken in the simulations shown in figure \ref{correction_simu}, the Debye layer thickness is 22 nm. In such conditions, the depletion in particles is perceptible up to 150 nm. Particles located in the depleted layer develop smaller excursions near the wall than far from it.  It follows that their apparent altitude, calculated from the nonlinear equation \ref{zappsimul}, is underestimated, while their apparent speed is overestimated. These particles thus stand apparently closer to the wall than they are, and move apparently faster than they do. This artefact gives rise to an apparent profile $V_a(z_a)$ that tends to level off as the wall is approached. \\
Another deviation between the apparent and expected profiles is visible at large distances, where the upper part of the profile is curved downwards. This effect originates in the fact that as we move farther from the wall, the particles become more and more unfocussed and consequently,  they seem farther than they are. A consequence of defocussing  is a slope reduction (down to 20\%), which gives rise to an overestimate of the in situ viscosity when determined from formula \ref{viscosity}. Note that these biases are small in amplitude. The consequence is that we do not need an outstanding accuracy on the calculation of the biases to correct the experimental data.

In practice, we use the function $V_a(z_a)$ to convert apparent velocity profiles into corrected profiles, which are expected to approach the \textquotedblleft true\textquotedblright velocity profiles.

\section{Experimental determination of $I_0$} \label{detI0}

\begin{figure}
\centering
\includegraphics[width=140mm]{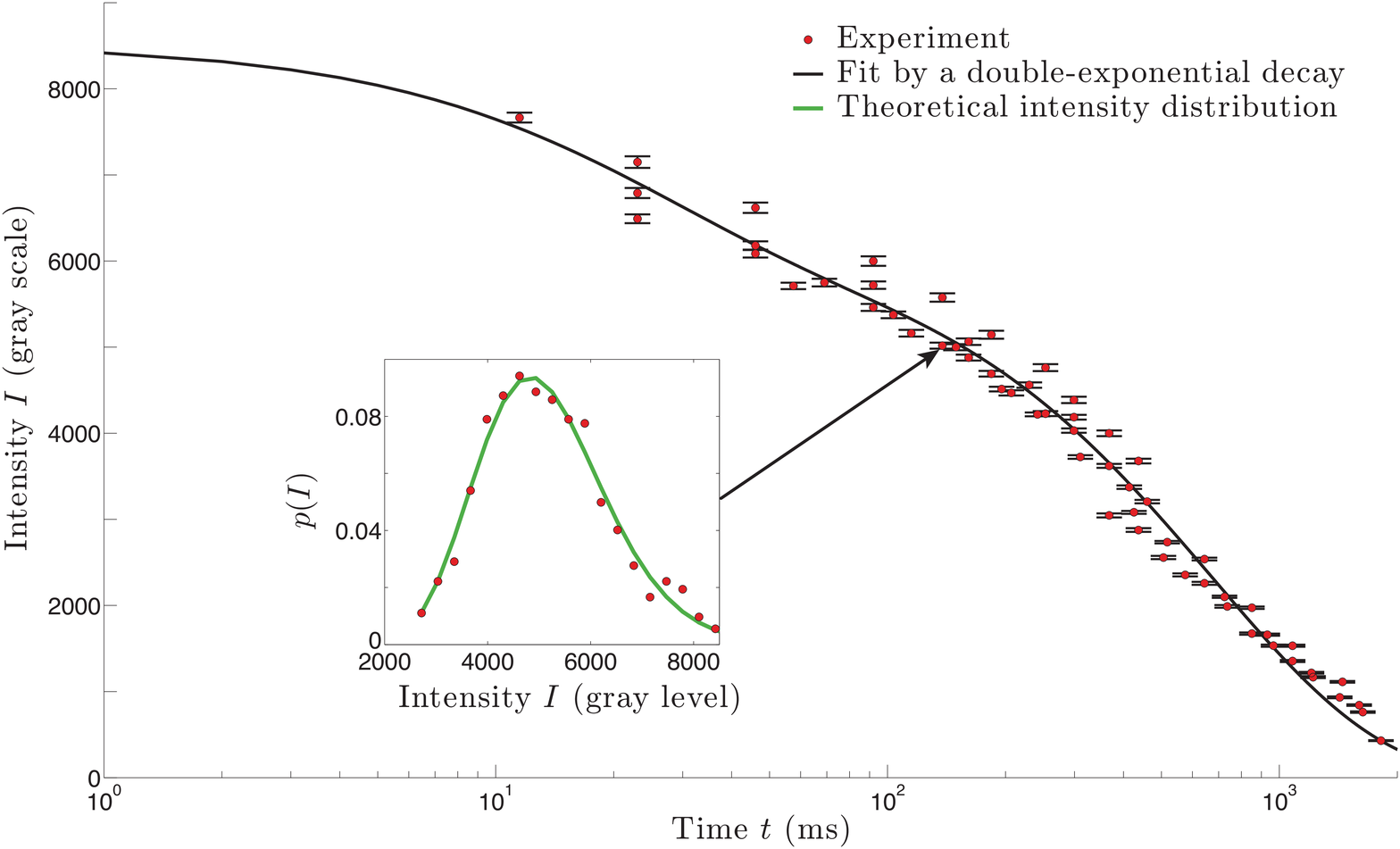}
\caption{Temporal decay of the intensity $I_B(t)$ emitted by particles physically adsorbed onto the wall, whose center are thereby approximately located at $z=r$. Experimental results are fitted by equation \ref{double-expon} which corresponds to double-exponential photo-bleaching kinetics \citep{song95,song97}, with $A$, $B$, $t_{1}$, and $t_{2}$ as free parameters. The fit parameters are $A=2283\pm 359$ grey values, $B=6300\pm 223$ grey values, $t_{1}=22.4\pm 9.5$ ms, $t_{2}=675\pm 49$ ms. The fit includes data obtained at $t < 1$ ms. In the insert, the red dots represent, for time $t=138$ ms, the intensity distribution emitted by the population of particles located at the surfaces; in the same insert, the green line is a fit based on equation \ref{fit-intensity}, from which $I_B(t)$ and $\sigma_{r}$ are determined.} 
\label{I0}
\end{figure}

$I_0$ is a crucial parameter that necessitates an accurate measurement. In order to measure this parameter, we analysed populations of particles physically adsorbed to the wall, thus located at $z=r$.  Adsorption of many particles onto the wall is achieved  by adding $0.05~$M NaCl to the working fluid, thus lowering the repelling electrostatic barrier between the particle and the wall. In our method, which is different from \cite{yoda10}, we first turn off the laser, maintain a flow, and wait for a few minutes, (i.e. in the black). This time is sufficient for the adsorption process to reach an equilibrium state, where all particles are immobilized. Then we turn the laser on, tune manually the objective position so as the particles lie in the focal plane of the TIRF objective. Throughout the procedure, a film is taken so that the instant at which the laser is switched on,  along with the time at which the first well focused image of the particles is captured are known with a 23 ms accuracy. No significant amount of particles gets adsorbed to the wall nor desorbs during the time the analysis is carried out. This hypothesis is checked directly on the movie. The inset of figure \ref{I0} shows an intensity distribution obtained 138 ms after the laser is switched on.

Such a distribution can be analysed theoretically. Assuming that the bleaching process is the same for all the particles, the intensity $I$ emitted by a particle of size $r$, located at $z=r$, and illuminated by the laser beam for a time $t$, is given by the formula :

\begin{equation}
I= I_B (t) \left( \frac{r}{r_0} \right)^3
\end{equation}

in which $I_B(t)$ is a function that characterizes the bleaching process, assumed to apply uniformly on the particle population, independently of their sizes. In this expression, we do not take into account the fact that the particles, being of different sizes, have their centers located at different distances from the wall and therefore are illuminated with slightly different intensities. The corresponding error is on the order of  $(\sigma_{r} r/p)^2$ i.e. $10^{-4}$. Assuming further a normal distribution for the particle size, with standard type deviation $\sigma_{r}$, one obtains the probability density function (pdf) of the intensity emitted by an ensemble of particles located at $z=r$ at time $t$:

\begin{equation}
p_W (I,t)=\frac{r_0}{3 I_B (t) \sigma_{r} \sqrt{2\upi}} \left(\frac{I_B (t)}{I}\right)^{2/3} \exp - \frac{ \left( \left( \frac{I}{I_B (t)} \right)^{1/3} -1 \right)^2} {2 \sigma_{r}^2} 
\label{fit-intensity}
\end{equation}

Our expression is slightly different from \cite{breuer06}, because we restored a factor $I^{-2/3}$ coming from a variable change, that was not taken into account. The inset of figure \ref{I0} shows that this distribution agrees well with the experiment. By comparing the theory and the experiments, we can extract $\sigma_{r}$  and $I_B (t)$  at each time. Regarding $\sigma_{r}$ , the values range between 8-9\% at all times, which is compatible with the data provided by the constructor (5\% for the size dispersity).  As mentioned in \cite{breuer06}, the standard deviation we obtain incorporates fluctuations in the particle size, number of fluorophores and quantum efficiencies. This may explain why it stands above the sole size dispersity stated by the constructor.

The evolution of  $I_B (t)$ with time is shown in figure \ref{I0}. We obtain an double-exponentially decreasing function, well represented by the following formula:
\begin{equation}
I_{B}(t)=Ae^{-t/t_{1}}+Be^{-t/t_{2}}
\label{double-expon}
\end{equation}

This behaviour reflects that, in the range of time we consider, two time constants are needed to describe the bleaching process, consistently with the literature \citep{song95,song97} and at variance with \cite{yoda08}. Figure \ref{I0} is important for the measurement we make, because it allows to check that the direct measurement of  $I_0$ (i.e that done at $t < 1$ ms, not represented on the figure), is in excellent agreement with the value $I_{B}(0)$ obtained with the fit. In practice, by repeating measurements at small times, the errors on the direct measurement of  $I_0$ are  estimated  on the order of $\pm$ 123 gray levels, which, translated in $z$-position by using formula \ref{evanescentfield}, leads to a $\pm$ 2 nm accuracy. 

 In \cite{yoda08, yoda10}  the photo bleaching kinetics of fluorescent particles was not analysed with such a degree of precision. This induced a substantial error on the average value of $I_{0}$, and therefore on the measurement of the slip length.

\section{Analysis of the intensity distributions and comparison with the Langevin simulations}

\begin{figure}
\centering
\includegraphics[width=120mm]{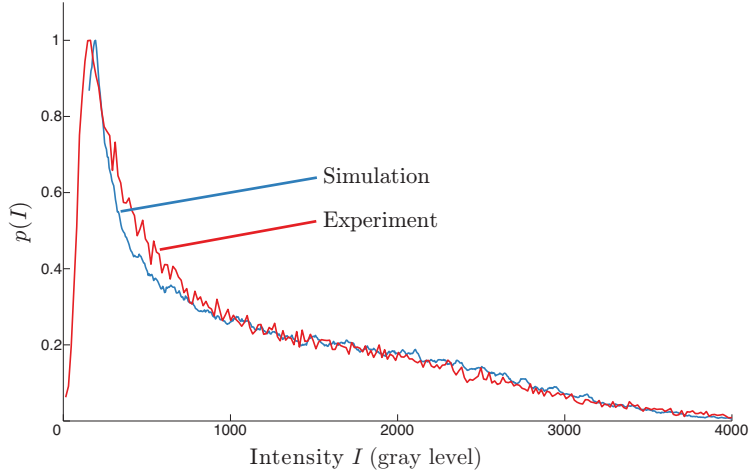}
\caption{Experimental (red line) and related simulated (blue line) intensity distributions in the case of 40w\% sucrose solution. Both situations correspond to an inlet pressure of 140 mbar. The simulation parameters are $\lambda_D = 22~$nm, $P_c = 348~$nm, $z_f = 300~$nm, $p = 124~$nm.} 
\label{intensitydistrib}
\end{figure}

The analysis of intensity distributions of populations of particles seeding the flow allows to confront our simulation to the experiment. The distributions shown in figure \ref{intensitydistrib} represents an experiment made with a sucrose solution. The distribution is strongly skewed. There are sharp peaks on the left side, and bumps on the right wing.

The left side of the distribution represents the contributions of the dimmest particles, i.e. those located far from the wall. If particles were uniformly distributed in space, the evanescent form of the intensity field would induce a hyperbolic function for the intensity distribution, which is consistent with the observed shapes of the central part of $p(I)$. Below a certain level, no particle is detected anymore, and the distribution level collapses. The largest intensity levels probe the closest distances to the wall. The abrupt decrease of $p(I)$ at high intensity levels is linked to the existence of depleted layers in the vicinity of the wall. These depleted layers result from the action of hindered diffusion and electrostatic repulsion. 

The simulation represents well the overall shape of the observed distributions. There are some discrepancies on the left side of the distribution, where the distribution of the dimmest particles seem to be imperfectly reproduced by the theory. We suggest that the discrepancy is due to the crude modeling of the conditions for which the particles cease to be observable by the camera. On the other hand, there are only small differences on the right part of the distribution, which correspond to the brightest particles, i.e those located closer to the wall. This suggests that the simplifications made in the theoretical representation of this region are acceptable. One weakness of the DLVO theory we used is that the particles are assumed pointwise, while they occupy a significant fraction of the Debye layer. Another weakness is that some parameters (such as the $\zeta$ potentials of the walls and particles) are poorly known. Still, from the viewpoint of the intensity distributions, these approximations seem acceptable.

As a whole, the agreement can be considered as satisfactory. This is an important observation, which supports the idea that  our Langevin model represents well the experiment, and thereby can provide reliable estimates of the existing biases. We also tested the robustness of the agreement with respect to changes in the parameters of the simulations. The lower range of intensities is sensitive to the size of the pack of particles we took for prescribing the initial conditions of the simulation. Likewise, in the upper range of intensity levels, a similar comment can be made regarding the sensitivity of the form of the distribution to the choice of the Debye layer thickness $\lambda_D$. Nonetheless, although quantitative differences may exist when the simulation parameters are not adequately chosen, the structures of the calculated distributions shown in figure \ref{intensitydistrib} are robust. Moreover,  changing the aforementioned parameters within a realistic range of values does not affect the velocity profiles outside the Debye layer. As a consequence, for the rest of the paper, we will concentrate ourselves on a range of altitudes lying between 200 nm and 600 nm for sucrose, 150 nm and 500 nm for water, for which the particles are sufficiently far from the wall to be insensitive to the detailed characteristics of the Debye layer, and more generally, for which the simulated velocity profiles are robust with respect to moderate changes of the simulation parameters.

\section{Analysis of the velocity distributions and comparison with the Langevin simulation}

\begin{figure}
\centering
\includegraphics[width=120mm]{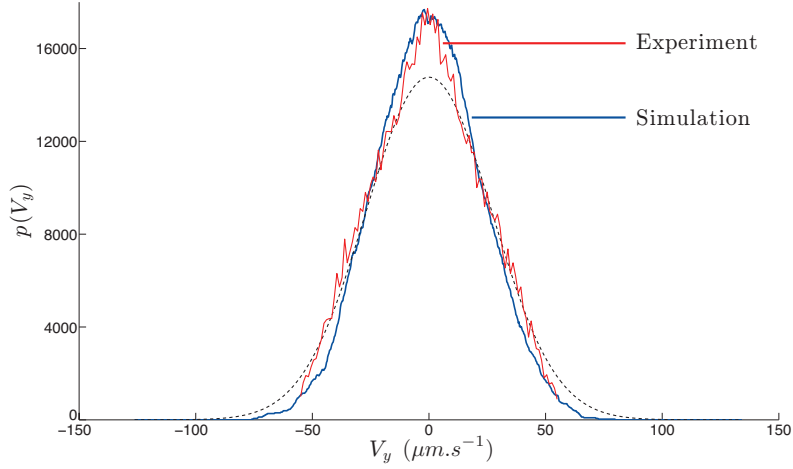}
\caption{Transverse velocity $V_y$ distribution. Red and blue lines correspond respectively to experimental and simulated data. The dotted line is the Gaussian distribution with standard deviation $\sigma_v = \left( \frac{2D}{\tau} \right) ^{\frac{1}{2}}$.} 
\label{distribV}
\end{figure}

A typical  transverse velocity distribution (i.e. along $y$), averaged across the $z$ interval 100 -- 600 nm, is shown in figure \ref{distribV}, and compared with the Langevin simulation.  Both agree well.

It is interesting to note that $V_y$ distribution is more peaked than a Gaussian curve, indicating that small  $V_y$ events are more probable than they would be for a Gaussian process. In fact, here, we look at the travelled distance along $y$ made by a large number of particles launched at the origin, after a time $\tau$. Within the range of time we considered, the process apparently keeps the memory of  the initial condition, giving rise to a cusp of the pdf around the origin. Since the distribution is not Gaussian, the diffusion constant associated to the $V_y$ distribution, is given by the following formula:
\begin{equation}
D = \frac{{\sigma_v}^2 \tau}{\alpha}
\end{equation}
with $\sigma_v$ being the standard deviation corresponding to the Gaussian distribution and $\alpha \approx 1.2$. Applying this formula in the experiments, we found diffusion constants in satisfactory agreement with the expectations. 

\section{Raw velocity profiles and raw viscosities}

\begin{figure}
\centering
\includegraphics[width=65mm]{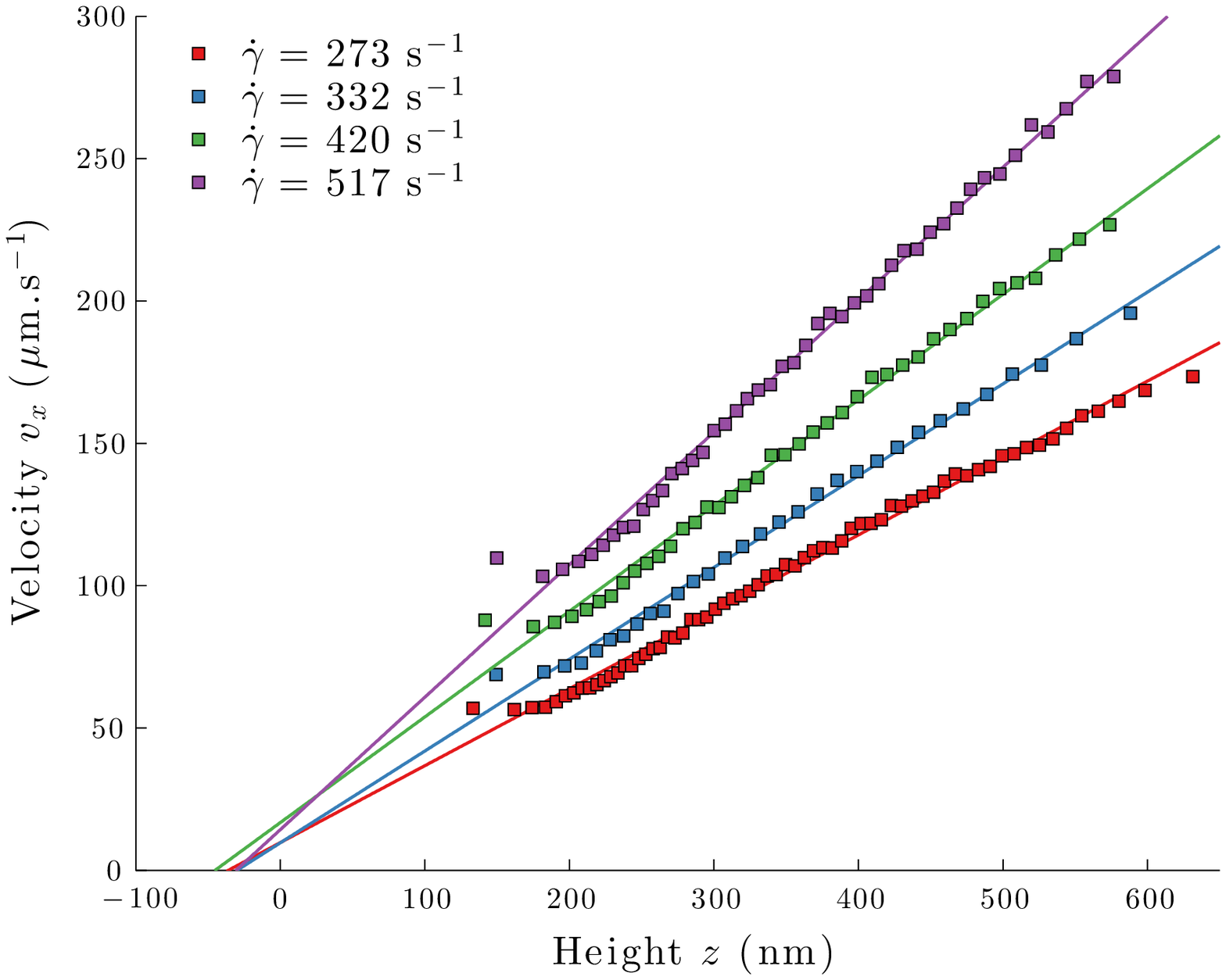} \includegraphics[width=65mm]{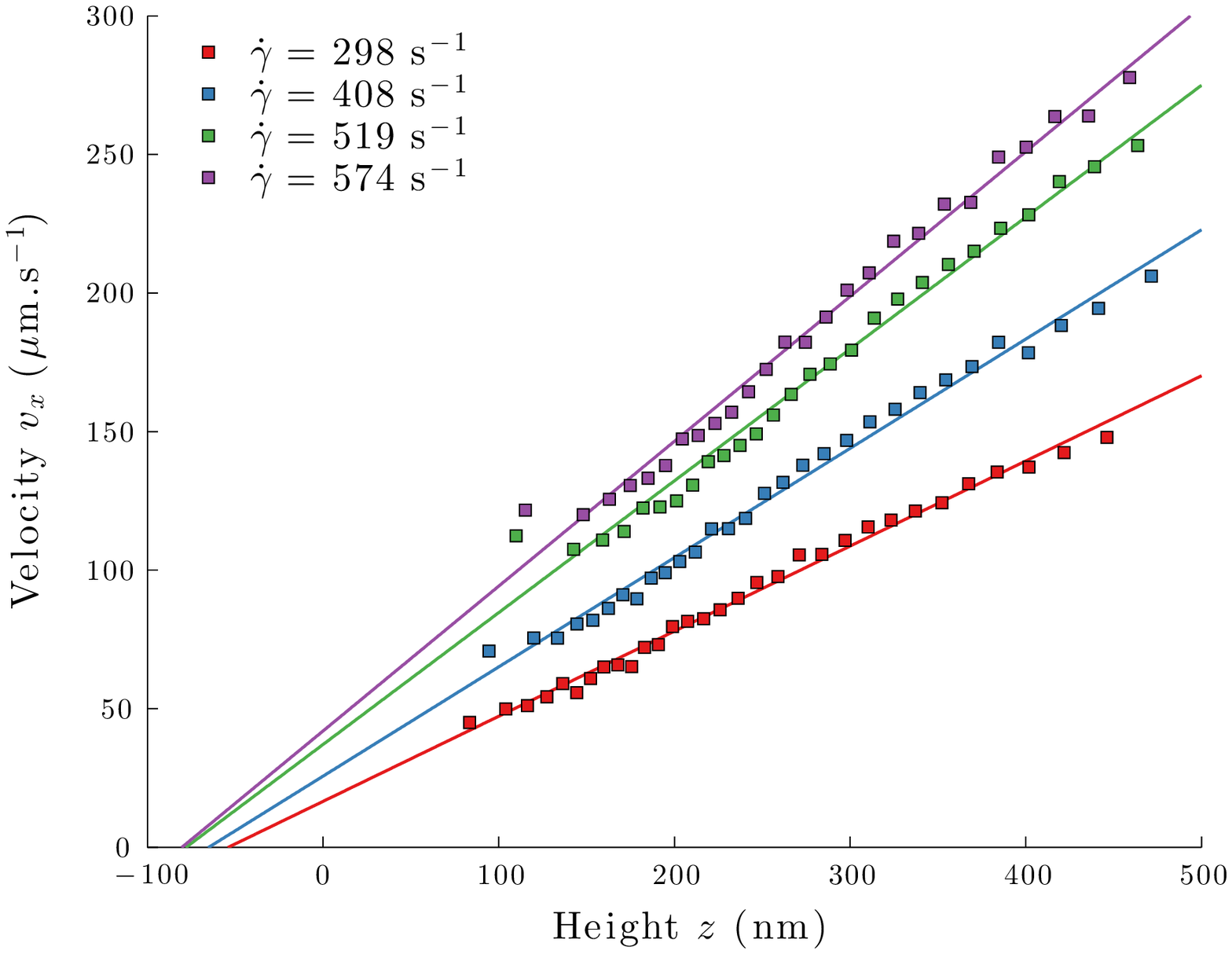} 
\caption{Uncorrected velocity profiles obtained with a 40w\% sucrose solution, over a hydrophilic surface (left), and with water over a hydrophobic surface (right), for different pressures. Linear fits are made between 200 and 600 nm for sucrose and 150 and 500 nm for water.} 
\label{raw}
\end{figure}

The raw velocity profiles  $V_x(z_a)$ are shown in figure \ref{raw} for the sucrose solution and water, pressures, as a function of the apparent altitude $z_a$. These profiles represent the raw data we obtained, i.e. those without corrections. 
Between 200 nm and 500 nm, the profiles are approximately linear in $z_a$. This is explained by the fact that on the scales we consider, the curvature of the Poiseuille profile is not visible and consequently, the profiles must be straight. 
 
Nonetheless, on looking in more detail, there is a systematic tendency to curve downwards above 500 nm. This is mainly due to defocussing, that weakens the intensity emitted by the particles, and in turn, increases their apparent altitude. Below 200 nm, the profiles tends to level off. This effect is due to the existence of a depletion zone close to the Debye layer, that favours a systematic underestimate in the measurement of the particle altitude and overestimated of speeds, as explained previously. According to the simulation, the levelling off effect is more pronounced when the Debye length is increased. This effect tends to vanish out as we move away from the walls, where the electrostatic force is out of scale, and particle concentration distribution tends to be homogeneous.

From the measurements of the profile slopes, in regions located between, typically, 200 and 600nm (where both electrostatic and defocussing biases are reduced), one may determine an apparent viscosity $\mu_a$, through the formula (see \ref{relationglobale}):

\begin{equation}
 \mu_a= \frac{h \Delta P}{2 \dot{\gamma}\ L}
\label{viscosity}
\end{equation}

\begin{figure}
\centering
\includegraphics[width=80mm]{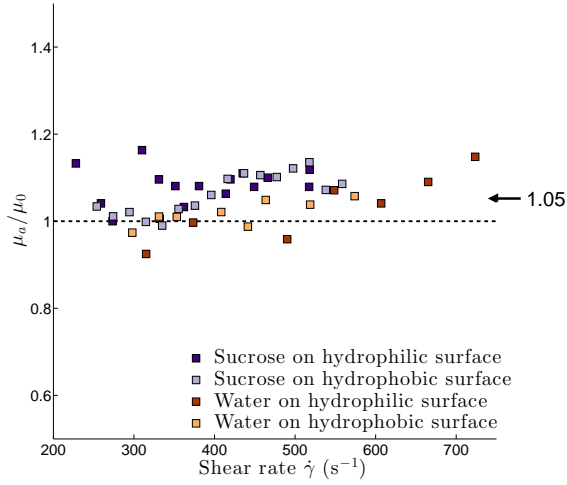}
\caption{Uncorrected measurements of the ratio between the  {\it in situ} apparent viscosity $\mu_a$ (see \ref{viscosity}) over the bulk viscosity $\mu_0$, determined independently with a rheometer. The mean deviation between the apparent and the bulk viscosity is 5\%.} 
\label{viscosityraw}
\end{figure}

 The measurements are shown in figure \ref{viscosityraw}. There are fluctuations (presumably linked to the $\pm~$50 nm uncertainties on the determination of the location of the focal plane) but it appears that viscosities systematically tend to stand above the expected values, by amount on the order of 5\%. The existence of this bias along with the order of magnitude of its importance, supports the analysis made in a precedent section.

\section{Corrected velocity profiles and corrected viscosities}

\begin{figure}
\centering
\includegraphics[width=70mm]{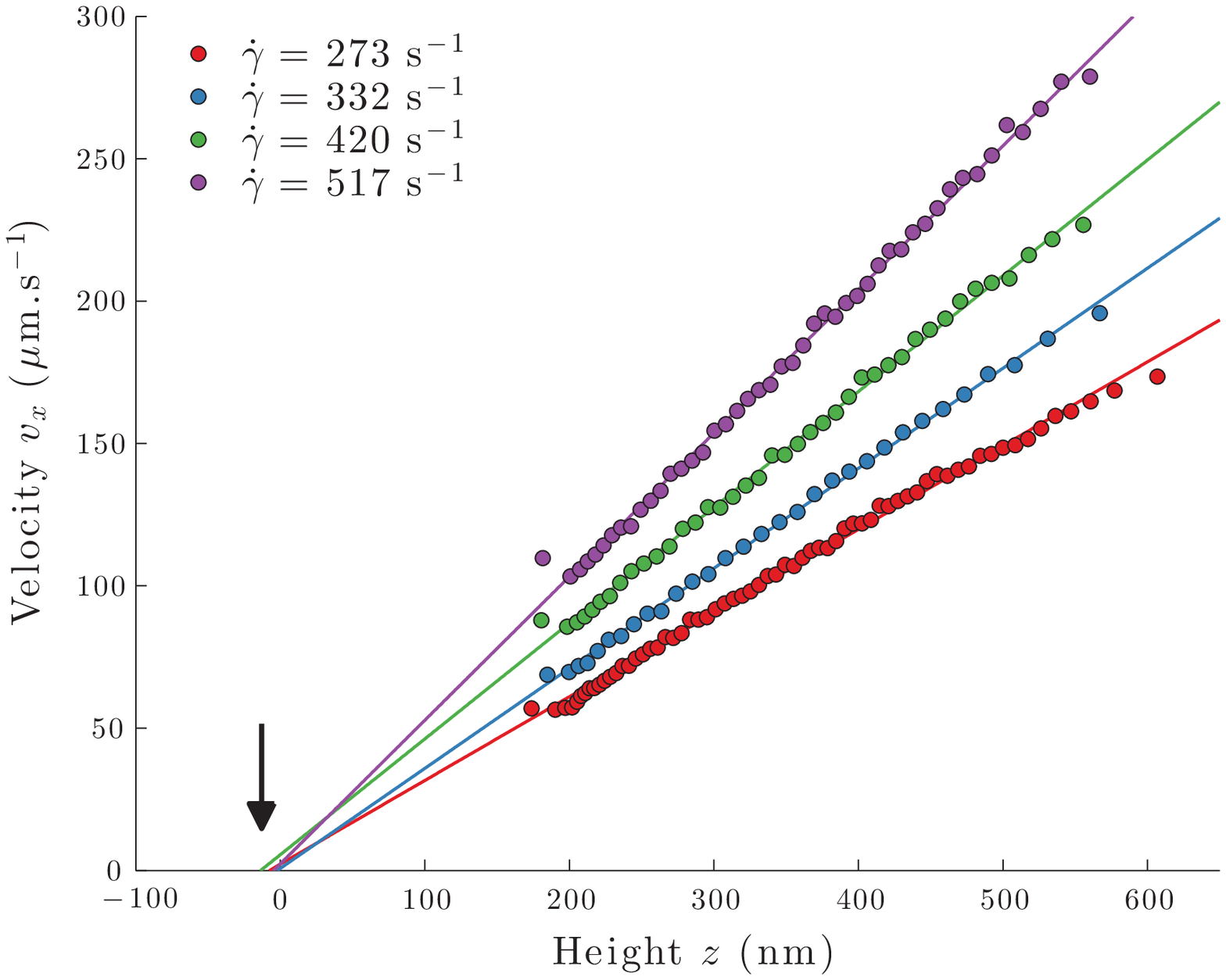}\includegraphics[width=70mm]{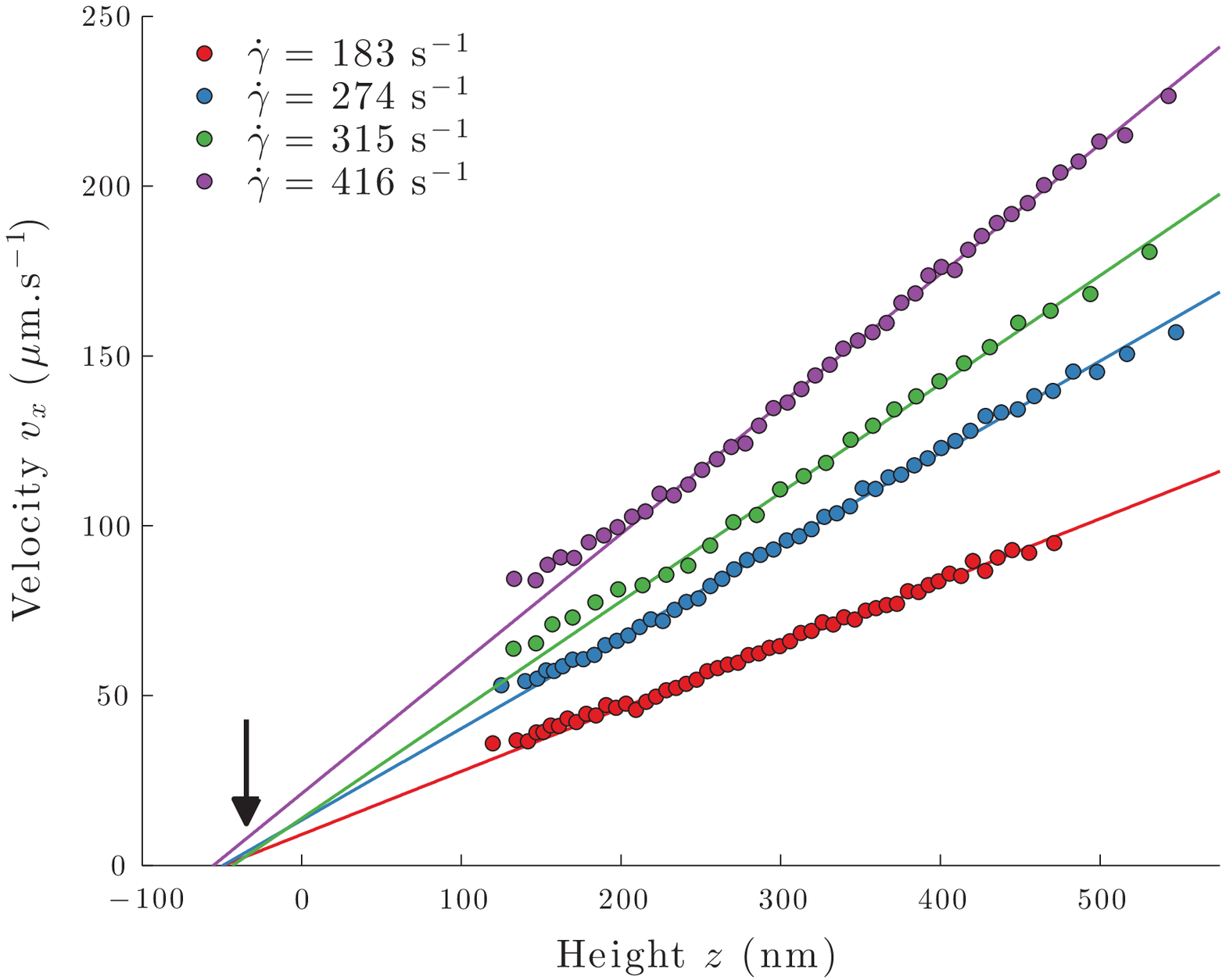}
\caption{Corrected velocity profiles obtained with sucrose solution 40w\%, on hydrophilic surface (left), and hydrophobic surface (right), for different pressures. The linear fit are calculated between 200 nm and 600nm. For hydrophilic walls, all profiles converge to a no-slip condition; with hydrophobic walls, they confirm the existence of slippage, with a slip length of 35 $ \pm~5$ nm.} 
\label{corrected1}
\end{figure}

\begin{figure}
\centering
\includegraphics[width=70mm]{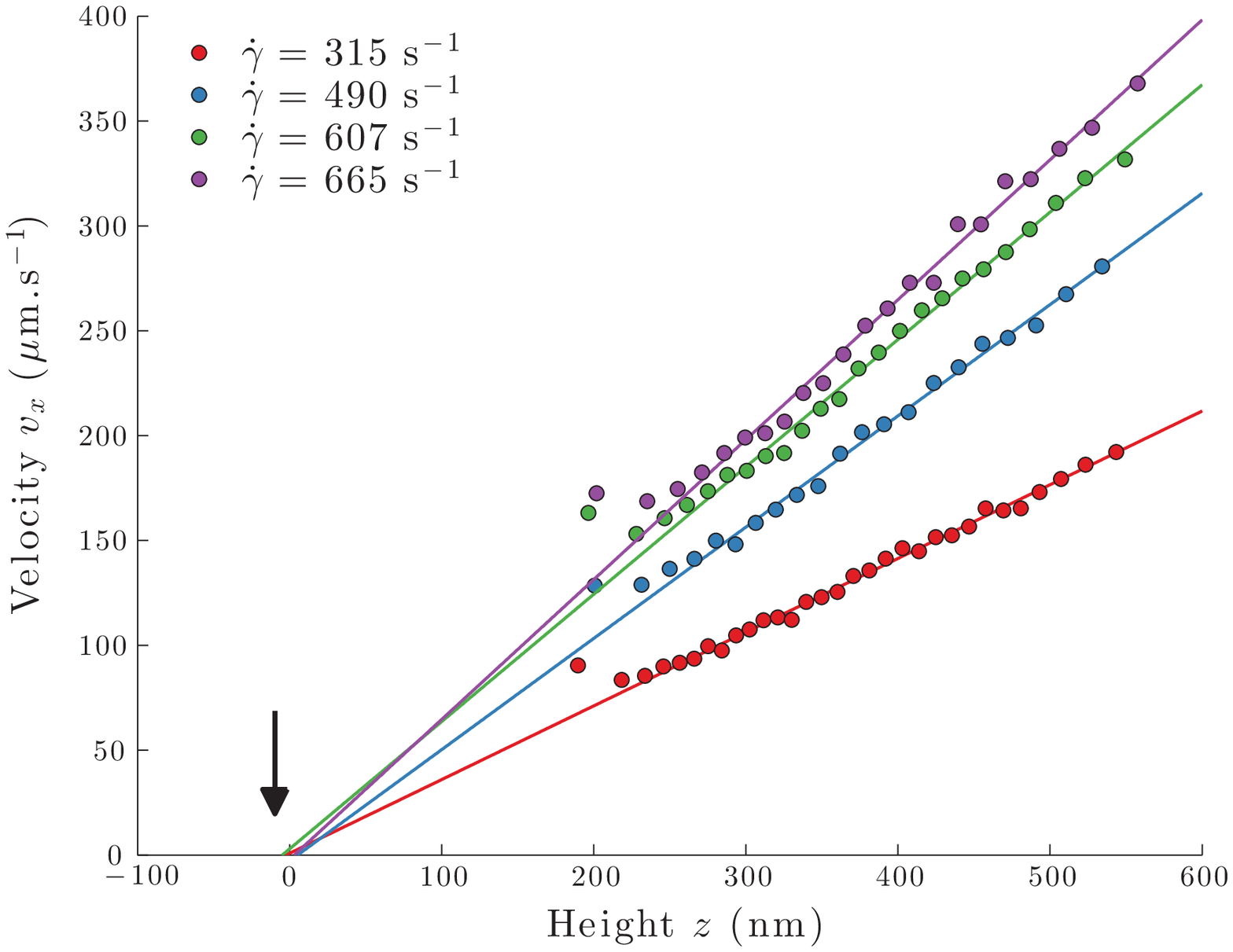}\includegraphics[width=70mm]{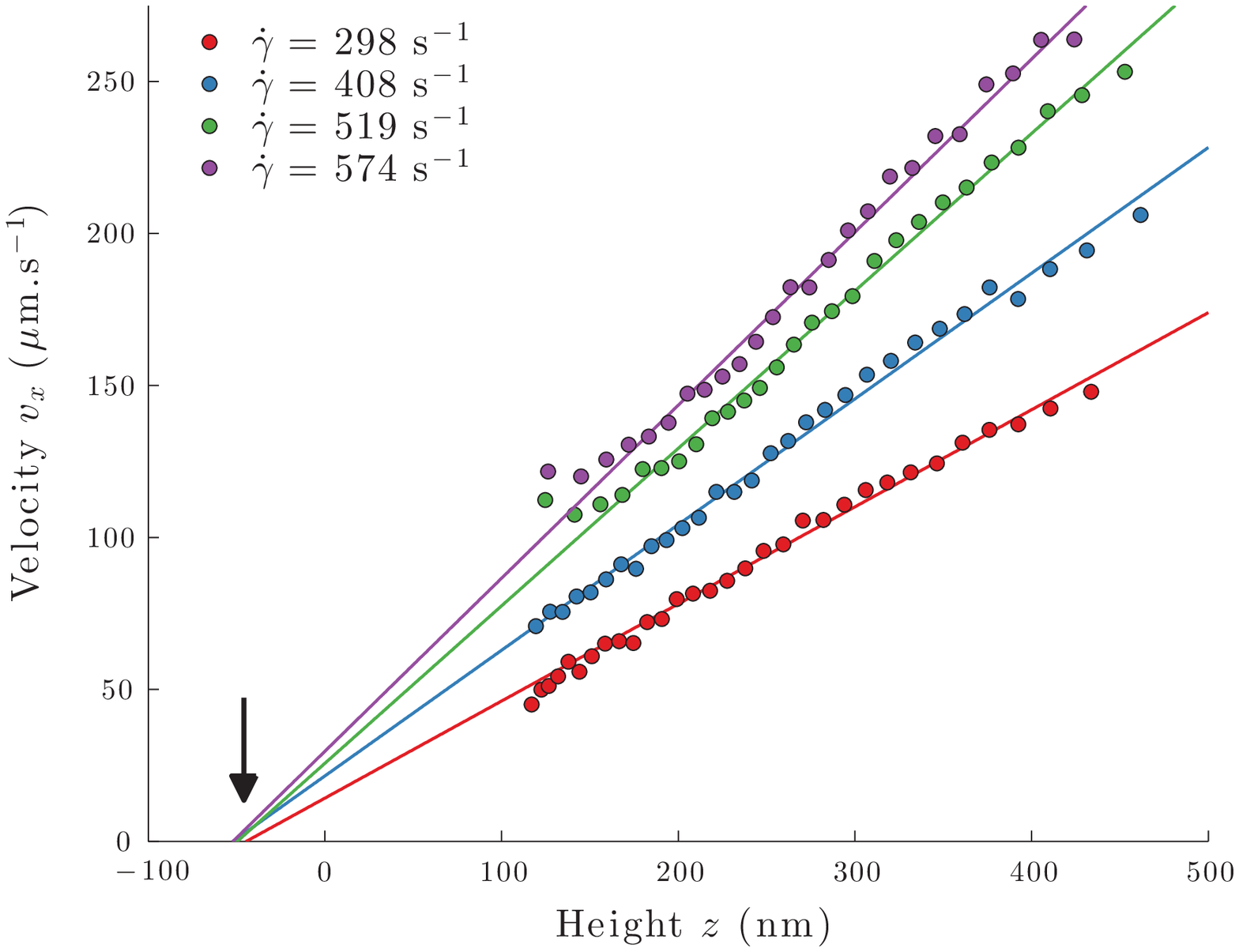}
\caption{Corrected velocity profiles obtained with water, on hydrophilic surface (left), and hydrophobic surface (right), for different pressures. The linear fits are calculated between 250 nm and 500 nm for the hydrophilic case and 150 and 500 nm for the hydrophobic one.} 
\label{corrected2}
\end{figure}

Figures \ref{corrected1} and \ref{corrected2} show the same set of data, but corrected, using formulas based on figure \ref{correction_simu}. 
The corrected data is close to the raw data, typical differences being on the order of 10\%.  
The corrected profiles are straight above about 200 nm, indicating that the tendency to curve down, as a result of defocussing, has been captured by the Langevin simulation and satisfactorily corrected. Moreover, the tendency to level off at small altitudes is less pronounced on the corrected than on the raw profiles.

\begin{figure}
\centering
\includegraphics[width=80mm]{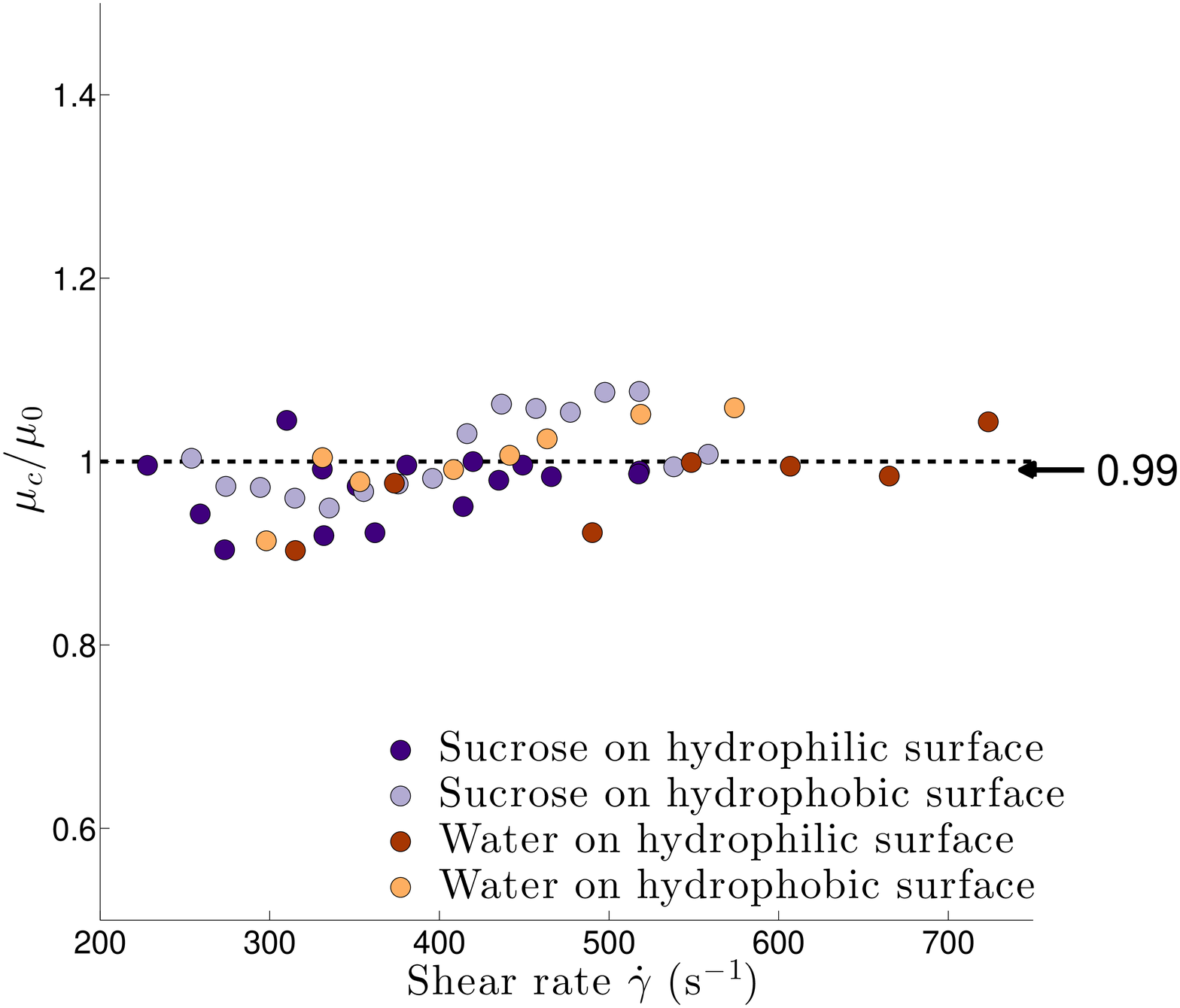}
\caption{Data showing corrected values of the {\it in situ} measured viscosity $\mu_c$ divided by the bulk viscosity $\mu_0$ determined independently. The graph includes data obtained with hydrophilic and hydrophobic walls.  The mean statistical deviation is 1\%.} 
\label{viscositycorr}
\end{figure}

The viscosities, obtained by using formula \ref{viscosity}, with shear rates estimated by fitting linearly the corrected profiles, are significantly closer to the expected values, the remaining discrepancy being on the order of 1\%, as shown on figure \ref{viscositycorr}. These results, associated to the remarks previously made on the shapes of the corrected profiles, support the validity of our method of correction.

\section{Slip length measurements}

We are now in position to carry out slip length measurements. As for the viscosity, slip lengths are obtained by fitting the corrected velocity profiles with straight lines, usually restricting ourselves to the 200 -- 600 nm range, and extrapolate them down to the $z$ axis. The (extrapolated) slip lengths are then measured as a function of the shear rate by localizing the intersections of this line with the horizontal axis, using the fit expressions.

\begin{figure}
\centering
\includegraphics[width=100mm]{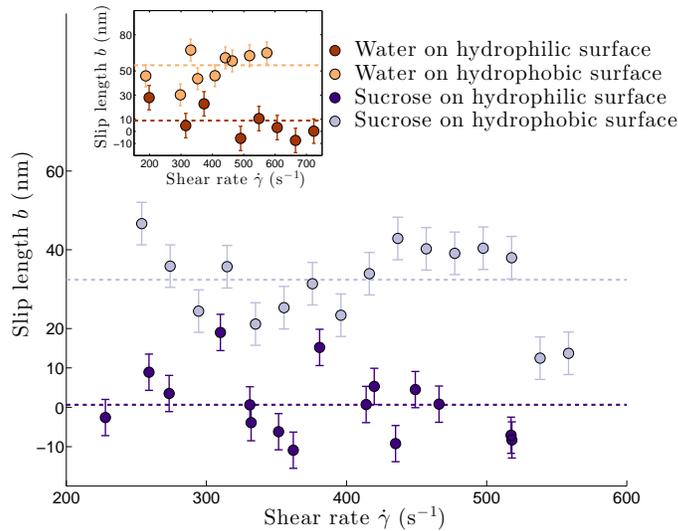}
\caption{Slip lengths measurements, obtained with the sucrose solution. The insert is the case of water. In all cases, one can easily detect a difference between hydrophilic (dark dots) and hydrophobic (light dots) walls.} 
\label{sliplength-fig}
\end{figure}

The results are plotted in figure \ref{sliplength-fig}, for hydrophilic, hydrophobic walls, using water and sucrose. For hydrophilic walls, one obtains, both for water and sucrose solutions, a slip length indistinguishable from zero. For hydrophobic walls, the slip lengths are clearly above the horizontal axis. 
It thus appears that the nature of the wetting property of the wall (i.e. whether it is hydrophilic or hydrophobic) is well captured by our slip length measurements, a result still not established firmly in the literature with TIRF based velocimetry, owing to the importance of the experimental uncertainties reported in the corresponding papers.

Taking the rule of 95\% confidence interval, and neglecting possible variations with the shear rate of the measured quantities - an hypothesis being acceptable owing to the small range of flow rate we explored - we  obtain 1 $\pm~$5 nm and 9 $\pm~$10 nm for hydrophilic walls (for the case of sucrose and water \textit{resp.}) and 32 $\pm~$5 nm and 55 $\pm~$9 nm for hydrophobic walls (again for sucrose and water respectively). This estimate of the errors does not include that made on $I_0$, which was found close to 2 nm (see Section \ref{detI0}), and which therefore can be neglected if we assume statistical independency. The hydrophilic data agrees well with the literature \citep{cottin-bizonne_boundary_2005}. Hydrophobic slip lengths stand somewhat above the published values, but stay in the right order of magnitude. With our present understanding of slippage phenomena, it is impossible to claim that there is an universal curve linking slip lengths to wetting angles, that would be totally independent of the detail of the surface physico-chemistry of the wall. Our data may suggest that the slip lengths lie in a region of the slip length-contact angle plot, rather than on a line, the spread being nurtured by slight differences in the surface physico-chemistry of the boundaries. 

\begin{table}
\begin{center}
\def~{\hphantom{0}}
\begin{tabular}{@{\extracolsep{4pt}}cccccc@{}}
            &       		&\multicolumn{2}{c}{Sucrose $40\%$w} & \multicolumn{2}{c}{Water}\\
\cline{3-4} \cline{5-6} 
\\
 \multicolumn{2}{c}{surface }          					& hydrophilic 	& hydrophobic  & hydrophilic  	& hydrophobic\\
\hline
$b$ 			&(nm)             		&	1		&      32       	&       9        	&      55    \\
$\Delta b$		&(nm)             		&      5       	&      5        	&       10       	&      9     \\
Error on viscosity &	$(\%)$  			&       2.7      	&     0.6         	&      2.8        	&     0.1      \\
\hline
\end{tabular}
\caption{Summary of our measurements made on slip lengths and viscosity errors}
\label{results}
 \end{center}
\end{table}

\section{Conclusion}

The first outcome of this paper is about the technique. We showed that nanoPTV technique, associated to Langevin stochastic calculations, achieves unprecedented accuracies ($\pm~$5 nm) on the determination of slip lengths, while delivering a local information, with an outstanding resolution of $\pm~$6 nm. This marks a substantial progress as compared to the state of the art, where error bars were such that it was difficult to draw out a firm conclusion on the existence of slippage over hydrophilic or hydrophobic walls.\\
 The reason why we succeeded to improve the accuracy of the technique is probably linked to a number of improvements concerning the determination of the wall position, the spatial resolution along with statistical conditions of measurement. It is remarkable that our Langevin simulations indicate that, with the methodology we took, systematic errors exist, but their amplitudes are small. This suggests that continuously tracking the particles represents, from the standpoint of the performances of the technique, an advantage in comparison with previous techniques. The sum of the improvements we made leads to  determine the velocity profiles more accurately and consequently determine the corresponding extrapolated slip lengths with a much better accuracy.\\
 We envision applications of our approach to non Newtonian flows, such as polymer solutions, microgels concentrated suspensions or living micelles, where the technique can readily be applied. These systems, in particular the polymer solutions, deserve a confirmation of slippage measurements made in the past with less sophisticated techniques.

Moreover, our work leads to interesting results concerning slip lengths. The slip lengths we found  for hydrophobic surfaces have the same order of magnitude as the literature. They also stand much above the numerical simulations. We thus confirm the unresolved discrepancy  between experiments and numerical simulations \citep{bocquet_charlaix10}. It challenges our understanding of the flow dynamics at nanometric distances from an interface, which in turn questions our ability to understand the physics of transport at solid/liquid interfaces in many systems, including natural ones. Owing to its importance, this unresolved issue deserves being addressed experimentally with different techniques, owing to the formidable difficulties that must be faced to perform  quantitative measurements in a range of scales that touch or pertain to the nanofluidic realm.

{\it Acknowledgments} The authors thank L. Bocquet, E. Charlaix, C. Cottin-Bizonne, T. Kitamori, D. Lohse, Z. Silber-Li, M. Tatoulian, and C. Ybert for  fruitful discussions and exchanges. Support from AEC, ESPCI, CNRS and UPMC is acknowledged.

%%%%%%%%%%%%%%%%%%%%%%%%%%%%%%%

\begin{appendix}

\section{Experiments dedicated to check the analysis of the defocussing effect}
\label{calibration}

In order to support the theoretical analysis of the effect of defocusing made in section \ref{langevinmod}, we carried out dedicated experiments. We moved the position of the focal plane $z_f$ by translating the TIRF objective with a piezo actuator (P-721~PIFOC nanopositioner controlled by E-662 LVPZT servo-amplifier, PI). At each position, we determined velocity profiles. Here, red and blue dots represent experimental data obtained with a 40w\% sucrose solution for an hydrophilic wall, for two different $z_f$,  200  $\pm$ 50 nm and 600  $\pm$ 50 nm. The results are fitted with an expression derived from equation \ref{intensityexpr} :

 \begin{equation}
V_x(z_{app}) = \dot{\gamma} \left[ z_{app} - p \log \left(1 +  \left(\frac{z_{app} - z_f}{P_{c}} \right) ^2 \right) \right]
\label{annexe}
\end{equation}

which captures the essence of the theoretical description made in section \ref{langevinmod}.  As shown in figure \ref{calibration}, we found that formula \ref{annexe} captures well the effect, within a range of altitudes where it is expected to apply, i.e., outside the Debye layer.
These experiments provided an experimental check of the theoretical description of the defocusing effect made in section \ref{langevinmod}.\\
In practice, we tuned $z_f$ between 200 and 360 nm in our Langevin model so as to minimize the curvature observed at long distances on the velocity profiles. The interval of  $z_f$  that is chosen is justified by the experimental procedure we took, which consists in adjusting the position of the objective to obtain a number of bright particles well in focus. Owing to the existence of a depletion in the Debye layer, the particles we focus on with this procedure are located, typically, between 200 and 400 nm from the wall.

\begin{figure}
\centering
\includegraphics[width=120mm]{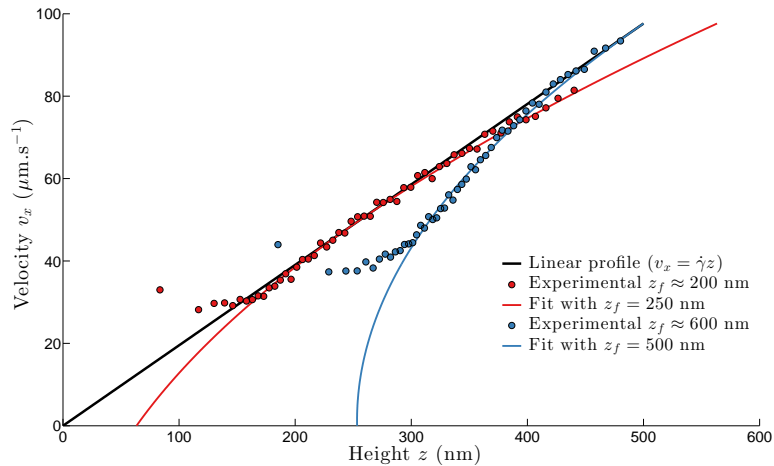}
\caption{Experiments showing the velocity profiles obtained at different positions of the TIRF objective, thus at different $z_f$. Red dots are obtained for  $z_f\approx200$ nm and blue dots correspond to $z_f\approx600$ nm. Full lines show good agreement between experimental data and equation \ref{annexe} within a range of altitudes where it is expected to apply, for the depth of field $a = 348$ nm of the objective.} 
\label{calibration}
\end{figure}

\end{appendix}

\bibliographystyle{jfm}

\bibliography{TIRFbib}

\end{document}